# Network-based Referral Mechanism in a Crowdfunding-based Marketing Pattern

Yongli LI [1, *], Zhi-Ping FAN, and Wei ZHANG

*1. School of Business Administration, Northeastern University, Shenyang 110169, P.R. China*
*2. School of Public Policy and Management, Central China Normal University, Wuhan 430079, P.R. China*

**Abstract:** Crowdfunding is gradually becoming a modern marketing pattern. By noting that the success of crowdfunding depends on network externalities, our research aims to utilize them to provide an applicable referral mechanism in a crowdfunding-based marketing pattern. In the context of network externalities, measuring the value of leading customers is chosen as the key to coping with the research problem by considering that leading customers take a critical stance in forming a referral network. Accordingly, two sequential-move game models (i.e., basic model and extended model) were established to measure the value of leading customers, and a skill of matrix transformation was adopted to solve the model by transforming a complicated multi-sequence game into a simple simultaneous-move game. Based on the defined value of leading customers, a network-based referral mechanism was proposed by exploring exactly how many awards are allocated along the customer sequence to encourage the leading customers' actions of successful recommendation and by demonstrating two general rules of awarding the referrals in our model setting. Moreover, the proposed solution approach helps deepen an understanding of the effect of the leading position, which is meaningful for designing more numerous referral approaches.

## 1. Introduction

With the widespread phenomenon of the sharing economy, crowdfunding is gradually moving beyond its traditional functions, such as funding artistic or creative projects, and becoming a modern marketing mode with the aim of selling various types of products or services, typically through online platforms (Brown et al. 2017). For example, several E-commerce platforms, such as JD (https://z.jd.com/sceneIndex._html) and Taobao (https://izhongchou.taobao.com/index.htm), contain a crowdfunding-based marketing pattern that is gradually being used to market numerous categories of products and services, such as clothes, fruit, wine, electronic products and even household services. Faced with this new marketing mode, the core problem we consider here is how to provide several marketing strategies in general and, in particular, an applicable referral mechanism for the process of crowdfunding-based marketing patterns.

---
[*] Corresponding Author. E-mail: ylli@mail.neu.edu.cn (Y. LI)

Generally, the crowdfunding process consists of three elements: the initiator who proposes the idea or project to be funded, the individuals or customers who support the idea, and a platform that brings the two parties together to conduct the crowdfunding (Mollick, 2014). If crowdfunding is understood to be a new marketing tool, it possesses several features. First, the crowdfunding initiator acts both as the sole producer and the seller of the crowdfunding project, although the producer and the seller are always different sections in the traditional supply chain. Second, crowdfunding has the function of market discovery; for example, if the crowdfunding is easily successful, it means that the targeted product or service is welcomed by the customers, and the initiator can in advance determine how many customers will buy the products or services, how much they will buy and who they are; if the crowdfunding is not successful until the deadline, the initiator can infer that the targeted product or service is not accepted by the market; thus, the initiator does not need to produce or to offer the good or the service. However, the traditional supply chain cannot perform as well as the crowdfunding because the traditional producer or seller cannot exactly know in advance whether the offered products or services will be accepted by customers or the extent of acceptance. Third, with the development of information technology, modern crowdfunding is much more dependent on online platforms; thus the word-of-mouth communication captured by local network externalities or the social learning embedded in social networks cannot be ignored, especially when potential consumers make their consumption decision and crowdfunding initiators offer their pricing and marketing strategies (Chu & Manchanda, 2016).

Accordingly, when we study and explore the referral mechanism for the progress of crowdfunding-based marketing patterns, we should pay close attention to the abovementioned three features. Specifically, in response to the first feature, we need only consider two types of actors in our model—the initiator and the customers—because the crowdfunding initiator acts as both producer and seller. Furthermore, the initiator can be regarded as the monopolist within the launched crowdfunding project, so that the established model features the monopoly structure. Please note that, although many crowdfunding projects can be launched almost at the same time in online platforms, the substitutability between these projects is always very low because the crowdfunding project often highlights its originality. Thus, one crowdfunding can be understood as a monopoly market for the targeted project especially during the launched period, which is often not long. Next, with regard to the second feature, the information known in advance can facilitate gaining the amount of participating customers (or, say, the market size), the customer sequence (or, say, the order of each customer who participates in the crowdfunding) and even the relationships of those who successfully recommend others to participate. Lastly, once the third feature is identified, the network externalities should be added into the analysis. In all, monopoly structure, customer sequence and network externalities are the three critical considerations in the process of solving the proposed research problem.

Since the monopoly structure has been discussed widely in the literature related to Economics and Management, we put it aside first while turning to analyse the other two considerations: network externalities and customer sequence. Here, we will focus on how to reflect and to depict them in our model and, further, to provide some intuitions about their potential relationships with possible market strategies. With regards to network externalities, in recent years, numerous researchers such as Anderson et al. (2013) and Wang & Wang (2017) have considered and studied them in online E-commercial platforms, especially in the context of monopoly structure, such as Crapis (2017), Ajorlou (2017) and Shin (2017). Based on these studies, it would be a feasible idea to utilize network externalities to improve marketing effectiveness (Bimpikis et al. 2016). For example, we can find the important nodes with highly positive network externalities and design mechanisms to give play to these nodes in enlarging markets and increasing profits (Li et al. 2014). Following this idea, we consider network externalities and their effects on customers' consumption decisions: the utility derived by a consumer from consuming a targeted good or service is significantly affected by his/her network peers who also consume that good or service. With regards to the customer sequence, it is a common phenomenon that some customers come into a crowdfunding project earlier than others. The earlier customers can potentially recommend the good or the service to their social network peers in a social media platform, and thus, these earlier customers are able to affect the marketing effects or even the profits earned by the crowdfunding initiator (or the monopolist, hereafter). In such a case, the earlier customers are called the leading customers (or leaders, hereafter) and the latter ones are called the following customers (or followers, hereafter). In fact, the idea of the Stackelberg game (Stackelberg, 1952) can be adopted in our model to reflect the mutual effects between leaders and followers.

Based on the three features of crowdfunding, the core idea of solving the research problem is to explore and measure the value of leading customers in the context of network externalities. Here, the value of leading customers is measured by the increased profits gained by the monopolist when some customers become the leaders from the known customer set. Once the value of the leading customers can be determined, we can calculate how much profit can be returned or awarded to the leading customers to raise their incentives to recommend more new customers. If so, the likelihood of successfully conducting a crowdfunding project can be increased. However, different customers can be located at different time points in the formed customer sequence, which means that some customers can be their earlier ones' followers and their later ones' leaders. Accordingly, one challenging problem is how to quantitatively allocate the awards to different customers with different time stamps along the customer sequence. Fortunately, the idea of measuring the value of the leading customer can also be adopted in this complicated case, where the contributions of different customers to the increased profits earned by the monopolist can be similarly determined. Accordingly, once the different contributions can be determined, the quantitative model of encouraging the leading customers to enlarge their recommendations can be proposed. In other words, the proposed approach of allocating the increased profits along the customer sequence can be adopted to award the action of successfully recommending new customers.

The remaining part of this section first reviews the related models in this field and then identifies our contributions. The following sections are organized as follows: Section 2 provides the sequential-move game model that highlights the three features: monopoly structure, network externalities and customer sequence; Section 3 presents the established model's consumption equilibrium and its optimal pricing strategy; Section 4 studies the relationship between our sequential-move game model and the known simultaneous-move game by way of matrix transformations and further discusses how to understand the influence of leading positions in the sequential-move game; Section 5 utilizes the results reported in Section 4 and focuses on measuring and illustrating the value of leading customers; Section 6 shows the application of our models and provides answers to the research problem; and lastly, Section 7 concludes.

### 1.1. Related work

Aside from the abovementioned background of crowdfunding, our study is also related to the scientific problem of how to design a customer's utility function to reflect the local network externality. Fortunately, Ballester et al. (2006) established a quadratic utility function that can reflect the peer effect of local network externality, and then Candogan (2012) and Bloch & Querou (2013) extended this utility function by adding the prices offered by the monopolist into the utility function. Because the designed quadratic utility function facilitates the mathematical analysis and is also able to reflect the law of diminishing marginal utility, it has been widely adopted in recent years by, for example, Zhou & Chen (2015), Li et al. (2015), Fainmesser & Galeotti (2015) and Zhou & Chen (2017). As mentioned above, crowdfunding features the monopoly and the network externalities, and thus, the work by Candogan (2012) and Bloch & Querou (2013) inspire the current study's approach to designing the utility function with local network externalities. Accordingly, this paper inherits the customer's utility functions designed by Candogan et al. (2012) and Bloch & Querou (2013).

Another important factor addressed here is the leading customer in the customer sequence, and we want to explore how the leader affects the followers' decisions and how the leading position affects the initiator's profits. Once the patterns of influence can be uncovered, we can provide marketing strategies by utilizing the leading customer's effects (Bass, 1969; Kerin et al., 1992). In fact, starting from the Stackelberg game model (Stackelberg, 1952), the effect of the leading customer has gradually been considered by an increasing number of researchers. For example, Markides & Sosa (2013) discussed the advantages and disadvantages of leading positions, and Klingebiel & Joseph (2016) revealed how firms make decisions on becoming the leading mover in innovation in the case of mobile phone technologies. Especially in the context of local network externalities, Zhou & Chen (2015) adopted the utility function designed by Ballester et al. (2006), analysed a two-stage game and then provided the approach to finding the key players in social networks. However, the work of Zhou & Chen (2015) does not consider the effect of the monopoly or generally the producers and sellers in the real market, but their idea is important for us in modelling the customer sequence. From this perspective, our work can be regarded as extending that of Zhou & Chen (2015) into a real market situation containing not only customers but also producers and sellers. In addition, our paper focuses on measuring the value of leading customers, aims to provide an applicable referral mechanism, and is thus different from Zhou & Chen (2015) and their extended work (Zhou & Chen, 2017).

### 1.2. Contributions

As crowdfunding is a relatively new phenomenon and a gradual marketing mode, the literature on crowdfunding has been growing.. Although these previous studies discuss many important problems in crowdfunding, such as the optimal pricing mechanism (Hu et al. 2014), the success likelihood (Parhankangas & Renko, 2017) and many others not mentioned, few studies focus on how to introduce the referral mechanism into a crowdfunding-based market pattern, let alone utilizing the network externalities to provide a network-based referral mechanism. More specifically, then, this paper not only analysed three features of modern crowdfunding—monopoly structure, network externalities and customer sequence—but also modelled them to offer several new management insights into how to increase the crowdfunding initiator's profits and how to raise the success likelihood of crowdfunding.

Second, this paper supplemented the theories of customer power by revealing a new customer power called by leading effect, which can be understood as a kind of demonstration effect. Traditionally, customer power is often understood as the ability of a customer to influence the decisions of a manufacturer (Brown et al., 1995; Goodman and Dion, 2001) and is divided into five sources: expert power, referent power, legitimate power, reward power and coercive power (Maloni & Benton, 2000). However, the measured value of the leading customer in this paper can be regarded as another customer power because the measured value can tell who can cause a larger profit increase when they become the leading customers. Accordingly, the contribution to uncovering another customer power will benefit the design of a network-based referral mechanism.

Note that one recent work (Zhou & Chen, 2017) that studied the optimal pricing with sequential consumption in networks found that the optimal sequence should be a chain structure. Although our model setting is quite similar to the cited work, our research aim is quite different. First, our work focuses on measuring the value of leading customers by solving the sequential-move game and further proposing an applicable network-based referral mechanism; however, the work of Zhou & Chen (2017) focused on finding the key players in a sequential-move game, uncovering the optimal sequence structure that causes the largest profit of the monopolist as well as comparing the different effects of symmetric and asymmetric social interactions on the pricing mechanism and the achieved profits. Second, although a similar solution technique was adopted in our work and that of Zhou & Chen (2017), our study focused on adopting the solution technique to link the sequential-move game with the equivalent simultaneous-move game and further deepening our understanding the effect of leading positions, whereas the work of Zhou & Chen (2017) did not focus on such problems. These differences, in our opinion, constitute one of our contributions. In fact, when we began this work and had finished most of it, the work of Zhou & Chen (2017) had not been published publicly in SSRN; thus, although some of the results in this paper, such as Lemma 2, are consistent with the work of Zhou & Chen (2017), the proof processes are quite different. However, despite these differences and although the main contributions are different, we cannot deny that the work of Candogan et al. (2012), Zhou & Chen (2015) and Zhou & Chen (2017) is foundational in this field and provides the important idea and basic framework for our study.

## 2. Model setup

### 2.1. Utility function with referral network

As we have explained, the crowdfunding-based marketing pattern consists of one monopolist who provides a divisible good and a finite set $N = \{1, 2, \cdots, n\}$ of consumers embedded in an information network that reflects the local network externality. Here, let the adjacency matrix $\textbf{\textit{IN}}$ represent the information network, and if customers $i$ and $j$ can share information with each other, the $ij$-entry and $ji$-entry of the matrix $\textbf{\textit{IN}}$ are assigned as 1; otherwise, they are assigned as 0. Based on the real existing network $\textbf{\textit{IN}}$, we further introduce and design a referral network $\textbf{\textit{R}}$ into the crowdfunding-based marketing pattern. Similar to the information network $\textbf{\textit{IN}}$, the $ij$-entry of $\textbf{\textit{R}}$ denoted as $r_{ij}$ also takes two values: $r_{ij} = 1$, which means customer $j$ successfully recommend customer $i$ to participate and $r_{ij} = 0$, which means not. Accordingly, $IN_{ij} = 1$ or $IN_{ji} = 1$ is the sufficient condition of $r_{ij} = 1$ by considering that sharing information is the precondition of recommendation; what's more, the referral network **R** is direct because the action of successful recommendation is naturally direct, whereas the information network **IN** is indirect. The established model will focus on the referral network **g** with the aim of how it influences the monopolist's profit and how to design it to increase the monopolist's profit.

Following the traditional utility form adopted in Candogan et al. (2012) and Bloch & Quérou (2013) as mentioned in the Related work section, we inherit the utility function of quadratic form, but we focus on the effect of our introduced referral network. In other words, although the network externality can originate from many aspects and takes many forms, this paper focuses on the effect of referral network. In this paper, we pay attention to ***two kinds of effects*** from the introduced referral network on the customer' utility: one is the influence from the referrers and the other comes from the successfully recommended customers. In other words, the utility of customer $i$ is not only affected by the customers who recommend the good to him/her but also affected by the customers who successfully accept his/her recommendation, and further, we allow the different influence strengths from the two directions. Let $\eta$ represent the different strengths, where $\eta > 1$ means the latter effect is stronger than the former and $1 > \eta > 0$ means the opposite. Accordingly, the specific utility function of customer $i$ is designed as

$$u_i(x_i; \mathbf{x}_{-i}, p_i) = \underbrace{\alpha_i x_i - \frac{1}{2}\beta_i x_i^2}_{\text{effect from individual preference}} + \underbrace{\left(\sum_{j=1}^{n} r_{ij} x_j\right) x_i}_{\text{influence from referrers}} + \underbrace{\left(\eta \sum_{j=1}^{n} r_{ji} x_j\right) x_i}_{\substack{\text{influence from the successfully}\\\text{recommended customers}}} \underbrace{-p_i x_i}_{\text{purchase cost}}, \qquad (1a)$$

where $x_i$ is the amount that consumer $i$ decides to purchase, $\mathbf{x}_{-i}$ is the consumption vector except consumer $i$, and $p_i$ is the price offered to consumer $i$ by the monopolist for one unit of the divisible good. Aside from the explained two effects in the middle of Equation (1a), two other important effects are also considered in the designed utility function. For the effect from individual preference, $\alpha_i$ and $\beta_i$ together represent the preference level of consumer $i$ on the good; furthermore, $\alpha_i$ measures the added utility by consuming one more unit of the good, and $\beta_i$ reflects the law of decreasing margin utility. For the purchase cost, the last term measures the cost of consumer $i$ by purchasing $x_i$ units of the good; in addition, the offer price $p_i$ can be different for different customers by noting that the monopolist enables price discrimination for different consumers according to their various purchasing time as well as positions in the referral network.

To facilitate expressing the results, we sum the two effects of referral network together to establish a new comprehensive network **g** of which the $ij$-entry is expressed as

$$g_{ij} = r_{ij} + \eta \cdot r_{ji}, \qquad (1b)$$

and then, the designed utility function in Equation (1a) can be simplified as

$$u_i(x_i; \mathbf{x}_{-i}, p_i) = \alpha_i x_i - \frac{1}{2}\beta_i x_i^2 + \left(\sum_{j=1}^{n} g_{ij} x_j\right) x_i - p_i x_i \, , \tag{1c}$$

whose form is identical with Candogan et al. (2012) and Bloch & Quérou (2013); however, the implication of the contained network is completely different. Moreover, the complicated form shown in Equation (1a) will also be useful to uncovering the implications of leading positions as well as to comparing the two effects (i.e., influence from referrers and influence from the successfully recommended customers) on increasing the monopolist's profit.

## 2.2. Profit function with differential pricing

In our model set, the monopolist is allowed to conduct complete price discrimination, and therefore customers can be offered different prices according to their preferences, purchasing sequence and network positions. Although obstacles can exist to hold back complete price discrimination in real markets, this assumption will also be useful in providing several foundational theoretical results. Next, the same production cost $c$ is assumed for each unit of the good, and then the monopolist faces the following profit function

$$\pi(x_i, p_i) = \sum_{i=1}^{n} (p_i - c) x_i \, . \tag{2}$$

Naturally, the monopolist will pursue the largest profits in the market by offering an optimal pricing strategy. In addition, we also assume that the monopolist can grasp the information network between customers, further enabling  control the formation of the referral network. In fact, it is not that surprising that the monopolist can grasp some information about customers, especially in the era of big data, because many social media platforms and numerous online shopping websites have been collecting customers' information day and night and can sell the relevant information to the monopolist.

## 2.3. Customer sequence embedded in the information network

It is a common phenomenon that some customers purchase a good earlier than others and then are likely to advertise that good along their local information network. To illustrate this phenomenon, we first divide all the customers into two categories: the leader category and the follower category, as the basic case. Further, to simplify their description, the first $m$ consumers numbered from 1 to $m$ are set as the leaders and the remaining $N-m$ consumers numbered from $m+1$ to $N$ are set as the followers in our basic model, and their sets are denoted as $a$ and $b$, respectively.

However, the ideal case is to divide the customers into two categories, as mentioned above. In real markets,

customers can comprise more than two categories in their purchasing sequence. To this end, the whole $N$ customers are divided into $k$ parts, where $k$ is an integer and $k \geq 2$. According to their purchasing order, the $k$ parts of customers are denoted as $t_1$, $t_2$, $\cdots$, $t_k$, respectively. As a result, these parts constitute the sequence set $T := \{t_1, t_2, \cdots, t_k\}$ that satisfies $t_i \cap t_j = \Phi$, $\forall i \neq j$ and $|\cup_{1 \leq i \leq k} t_i| = n$.

The two kinds of participants—the monopolist and the customers with purchasing sequence—form a game in which the monopolist decides the price offered to each customer and the customers decide their consumptions. In our model, when the leader parts make their consumption decision, they will consider the responses of the monopolist and the follower parts; and when the follower parts make their consumption decision, the leaders' consumption is known information for them and they will only consider the response of the monopolist. To clarify, we first discuss the basic case and then extend it to the complicated case. Note that the solution of these cases establish a theoretical foundation to measure the value of leading customers and, further, to propose an applicable referral mechanism in the context of network externality in general and crowdfunding-based marketing patterns in particular.

## 3. Basic model, solutions and discussions

### 3.1. Solution process for the basic model

Note that the basic model considers that all the customers are only divided into two categories: the leader set and the follower set. Hence, following the work of Candogan et al. (2012), Bloch & Quérou (2013) and Zhou & Chen (2015), the whole solution process for the basic model consists of three steps.

***Step 1***: **Solving the followers' consumption strategy**.

Let $\mathbf{x}^a := (x_1^a, x_2^a, ..., x_m^a)^{\mathrm{T}}$ and $\mathbf{x}^b := (x_{m+1}^b, x_{m+2}^b, ..., x_N^b)^{\mathrm{T}}$ denote the leaders' and the followers' consumption vectors, respectively. When the followers make their consumption strategy, $\mathbf{x}^a$ is regarded as known for them by considering that the followers can directly observe the leaders' announced order in the online platform. Accordingly, the followers' consumption strategy can be achieved by maximizing their utilities, given the prices offered by the monopolist and the leaders' consumption, i.e.,

$$x_j^b = \arg\max_{y_j^b \in [0, \infty)} u_j(y_j^b, \mathbf{x}_{-j}^b \mid \mathbf{x}^a, p_j), \ (j = m+1, m+2, ..., N). \tag{3}$$

As a result, the solution of the above formula can lead to the following optimal response function:

$$\mathbf{x}^b = f_1(\mathbf{x}^a, \mathbf{p}), \tag{4}$$

where the vector $\mathbf{p}$ consists of the offered price of each customer and $f_1$ is a quasi-linear function by considering the quadratic form of the utility function. The optimal response function allows $\mathbf{x}^b$ to be explicitly expressed by the variables $\mathbf{x}^a$ and $\mathbf{p}$, which lay the foundation for the next stages and further solutions.

*Step 2*: **Solving the leaders' consumption strategy**.

By substituting Equation (4) into the leaders' utility function $u_i(\mathbf{x}^a, \mathbf{x}^b, \mathbf{p})$ $(i = 1, 2, ..., m)$, the leaders' utilities can be transferred as the function of $\mathbf{x}^a$ and $\mathbf{p}$, i.e., $v_i(\mathbf{x}^a, \mathbf{p}) := u_i(\mathbf{x}^a, f_1(\mathbf{x}^a, \mathbf{p}), \mathbf{p})$. Accordingly, the leaders' consumption strategy can be achieved by maximizing $v_i(\mathbf{x}^a, \mathbf{p})$ with the following specific expression:

$$x_i^a = \underset{y_i^a \in [0, \infty)}{\arg\max} v_i(y_i^a, \mathbf{x}_{-i}^a, \mathbf{p}), \ (i = 1, 2, ..., m). \tag{5}$$

Similar to the solution process of *Step 1*, Equation (5) also leads to the optimal response function of the leaders' consumptions to the offered prices, i.e.,

$$\mathbf{x}^a = f_2(\mathbf{p}), \tag{6}$$

where $f_2$ is also a quasi-linear function following the same reasons of $f_1$. Then, by substituting Equation (6) into Equation (4), $\mathbf{x}^b$ can further be simplified as the function of the offered prices, specifically

$$\mathbf{x}^b := f_3(\mathbf{p}) = f_1(f_2(\mathbf{p}), \mathbf{p}). \tag{7}$$

*Step 3*: **The monopolist's pricing strategy**.

Recalling Equation (2), the profit function of the monopolist can further be expressed as the function of the price vector $\mathbf{p}$ by substituting Equations (6) and (7), i.e.,

$$b(\mathbf{p}) = \mathbf{p}^{\mathrm{T}} \begin{pmatrix} \mathbf{x}^a \\ \mathbf{x}^b \end{pmatrix} = \mathbf{p}^{\mathrm{T}} \begin{pmatrix} f_2(\mathbf{p}) \\ f_3(\mathbf{p}) \end{pmatrix}, \tag{8}$$

and therefore, the optimal pricing strategy of the monopolist can be immediately achieved by maximizing the profit $b(\mathbf{p})$ in Equation (8); i.e.,

$$\mathbf{p}^* = \arg\max b(\mathbf{p}). \tag{9}$$

Next, the optimal consumption strategies $\mathbf{x}^a *$ and $\mathbf{x}^b *$ can be achieved by substituting $\mathbf{p} *$ into Equations (6) and (7), respectively. Note that the price in the optimal pricing strategy is allowed to be less than $c$ or even less than 0, and thus, no constraints are set for the targeted pricing strategy, because we allow the monopolist to pay some consumers if the total profit can be raised.

### 3.2. Assumptions and solutions of basic model

The comprehensive network $\mathbf{g}$ defined in **Section 2.1** is divided into four parts by the two categories of

customers:

$$\mathbf{g} = \begin{bmatrix} \mathbf{g}^{aa} & \mathbf{g}^{ab} \\ \mathbf{g}^{ba} & \mathbf{g}^{bb} \end{bmatrix}. \tag{10}$$

Based on this expression, Assumptions 1 and 2 are discussed first to guarantee reasonable solutions.

[**Assumption 1**] Each customer's preference parameter $\beta_i$ is assumed to be large enough to guarantee that the following two matrixes $\Lambda_{\boldsymbol{\beta}}^b - \mathbf{g}^{bb}$ and $\Lambda_{\boldsymbol{\beta}}^a - \mathbf{g}^{aa} - \mathbf{g}^{ab}\mathbf{A}^b\mathbf{g}^{ba} - diag(\mathbf{g}^{ab}\mathbf{A}^b\mathbf{g}^{ba})$ are invertible, where $\Lambda_{\boldsymbol{\beta}}^a := diag(\beta_1, \beta_2, \cdots, \beta_m)$, $\Lambda_{\boldsymbol{\beta}}^b := diag(\beta_{m+1}, \beta_{m+2}, \cdots, \beta_N)$, $\mathbf{A}^b := [\Lambda_{\boldsymbol{\beta}}^b - \mathbf{g}^{bb}]^{-1}$, and $diag(\mathbf{g}^{ab}\mathbf{A}^b\mathbf{g}^{ba})$ denotes the diagonal matrix consisting of the diagonal elements of $\mathbf{g}^{ab}\mathbf{A}^b\mathbf{g}^{ba}$. □

**Assumption 1** guarantees that the law of diminishing marginal utility plays a dominant role even if the network effects are considered. Generally, this assumption is often met in the real market because the customer's decision is often rational, or at least to some extent. Otherwise, if the influence of the network effect plays a dominant role, some customers will buy the goods as much as possible. In most common cases, the extreme phenomenon seldom appears in the real market, and thus, **Assumption 1** guarantees a common case in real markets.

[**Assumption 2**] $\alpha_i > c$ holds for all the customers. □

**Assumption 2** guarantees that the achieved optimal consumption vector is no less than zero by recalling Equation (13). In fact, this assumption means that all the concerned customers in the market truly need the good offered by the monopolist. It is true that some customers do not need the good; however, these customers do not come into our consideration, as they do not participate in the crowdfunding project.

Based on the two assumptions and the three steps in Section 3.1, the results of the basic model are achieved below.

[**Result 1**] (**Solutions of basic model**). When the customers are divided into two categories—leader set and follower set—the optimal consumption vector $\mathbf{x}_{se}$ of the customers is

$$\mathbf{x}_{se} = [\mathbf{M}^{-1} + \mathbf{M}^{-T}]^{-1}(\boldsymbol{\alpha} - c\mathbf{1}), \tag{11}$$

the optimal offered price vector $\mathbf{p}_{se}$ is

$$\mathbf{p}_{se} = \boldsymbol{\alpha} - \mathbf{M}^{-1}[\mathbf{M}^{-1} + \mathbf{M}^{-T}]^{-1}(\boldsymbol{\alpha} - c\mathbf{1}), \tag{12}$$

and, accordingly, the largest profit $\pi_{se}$ gained by the monopolist is

$$\pi_{se} = \frac{(\boldsymbol{\alpha} - c\mathbf{1})^{\mathrm{T}}[\mathbf{M}^{-1} + \mathbf{M}^{-T}]^{-1}(\boldsymbol{\alpha} - c\mathbf{1})}{2}. \tag{13}$$

Here, matrix $\mathbf{M}$ has the specific form:

$$\mathbf{M} = \begin{bmatrix} \mathbf{B}_{se}^a & \mathbf{B}_{se}^a \mathbf{g}^{ab} \mathbf{A}^b \\ \mathbf{A}^b \mathbf{g}^{ba} \mathbf{B}_{se}^a & \mathbf{A}^b + \mathbf{A}^b \mathbf{g}^{ba} \mathbf{B}_{se}^a \mathbf{g}^{ab} \mathbf{A}^b \end{bmatrix}. \tag{14}$$

***Proof***. See **Appendix A** for details. □

It is worth noting that the achieved optimal solutions are all highlighted by the subscript "*se*" to differentiate between the results of the *simultaneous-move game*.

### *3.3. Relationship with the simultaneous-move game*

The basic model belongs to a *sequential-move game*, and thus, an interesting question we want to explore is whether a *sequential-move game* can be transformed into an equivalent *simultaneous-move game*, where the "equivalent" means the two kinds of game shares the identical results. If the answer is yes, we can further cope with the extended model with a complicated customer sequence by considering that the equivalent *simultaneous-move game* is comparatively easy to solve.

To this end, we first recall the work of Candogan et al. (2012) on the *simultaneous-move game* under a similar model setting to ours, except that all the customers in Candogan et al. (2012) make their consumption decisions simultaneously. To distinguish the *sequential-move game*, we use the subscript "*si*" to denote the *simultaneous-move game*, and the relevant results of Candogan et al. (2012) are listed below.

[**Result 2**] (The *simultaneous-move game* dealt with by Candogan et al. (2012)). The optimal consumption vector $\mathbf{x}_{si}$ of all the customers is

$$\mathbf{x}_{si} = [\mathbf{A}^{-1} + \mathbf{A}^{-\mathrm{T}}]^{-1} (\boldsymbol{\alpha} - c\mathbf{1}), \tag{15}$$

where $\mathbf{A} := [\Lambda_{\boldsymbol{\beta}} - \mathbf{g}]^{-1}$ and $\Lambda_{\boldsymbol{\beta}} := diag(\beta_1, \beta_2, \cdots, \beta_N)$. The optimal offered price vector $\mathbf{p}_{si}$ is

$$\mathbf{p}_{si} = \boldsymbol{\alpha} - \mathbf{A}^{-1} [\mathbf{A}^{-1} + \mathbf{A}^{-\mathrm{T}}]^{-1} (\boldsymbol{\alpha} - c\mathbf{1}), \tag{16}$$

and accordingly, the largest profit $\pi_{si}$ gained by the monopolist is

$$\pi_{si} = \frac{(\boldsymbol{\alpha} - c\mathbf{1})^{\mathrm{T}} [\mathbf{A}^{-1} + \mathbf{A}^{-\mathrm{T}}]^{-1} (\boldsymbol{\alpha} - c\mathbf{1})}{2}. \tag{17}$$

□

By comparing Result 1 and Result 2, we find that the two results share similar mathematical expressions except for matrixes **M** and **A**. Accordingly, the key to linking the two kinds of games is to uncover the relation between the two matrixes **M** and **A**. To answer this question, matrix **A** is further re-expressed as

$$\mathbf{A} = [\Lambda_{\boldsymbol{\beta}} - \mathbf{g}]^{-1} = \begin{bmatrix} \Lambda_{\boldsymbol{\beta}}^a - \mathbf{g}^{aa} & -\mathbf{g}^{ab} \\ -\mathbf{g}^{ba} & \Lambda_{\boldsymbol{\beta}}^b - \mathbf{g}^{bb} \end{bmatrix}^{-1} = \begin{bmatrix} \mathbf{B}_{si}^a & \mathbf{B}_{si}^a \mathbf{g}^{ab} \mathbf{A}^b \\ \mathbf{A}^b \mathbf{g}^{ba} \mathbf{B}_{si}^a & \mathbf{A}^b + \mathbf{A}^b \mathbf{g}^{ba} \mathbf{B}_{si}^a \mathbf{g}^{ab} \mathbf{A}^b \end{bmatrix}, \tag{18}$$

where $\mathbf{B}_{si}^a := \left[ \Lambda_{\boldsymbol{\beta}}^a - \mathbf{g}^{aa} - \mathbf{g}^{ab} \mathbf{A}^b \mathbf{g}^{ba} \right]^{-1}$. Thus, the two matrixes **M** and **A** share a similar structure, except for the difference between $\mathbf{B}_{se}^a$ and $\mathbf{B}_{si}^a$. Now, by focusing on the difference between $\mathbf{B}_{se}^a$ and $\mathbf{B}_{si}^a$, an additional term $diag(\mathbf{g}^{ab}\mathbf{A}^b\mathbf{g}^{ba})$ is found in $\mathbf{B}_{se}^a$. The subtle difference inspires us to transform the matrix $\mathbf{g}$ to equate the *sequential-move game* with the *simultaneous-move game* in our model setting. As a result, let $\mathbf{g}'$ be the transformed matrix from $\mathbf{g}$, and its specific expression is provided in **Lemma 1**.

[**Lemma 1**] (The specific expression of $\mathbf{g}'$). Let $\mathbf{g}^{aa}{}'=\mathbf{g}^{aa}+diag(\mathbf{g}^{ab}\mathbf{A}^b\mathbf{g}^{ba})$ and the remaining three parts in $\mathbf{g}'$ remain identical to the corresponding parts in $\mathbf{g}$, then the optimal consumption, the optimal pricing strategy and the optimal profit achieved from the simultaneous-move game under $\mathbf{g}'$ are equal to those from the sequential-move game under $\mathbf{g}$.

***Proof***. See **Appendix B** for details. □

As shown in **Lemma 1**, when customer $i$ is a leading customer, the effect of his/her leading position in the *sequential-move game* can be reflected by adding some value to $g_{ii}$ in the corresponding *simultaneous-move game*. Note that before the matrix transformation, $g_{ii}=0$ according to our model setting. **Lemma 1** is important in understanding the relationship between the *simultaneous-move game* with the *sequential-move game* in our model setting. In addition, the matrix $\mathbf{M}$ obtained in the process of the *sequential-move game* is not convenient for mathematical expression and theoretical analysis, and therefore it is meaningful to find an equivalent *simultaneous-move game* for ease of mathematical analysis. Lastly, the idea of matrix transformation is useful not only for the case in which the customers are divided into two categories but also for the case in which the customers are divided into multiple categories according to their purchasing sequences.

### 3.4. Discussions

Unlike the results reported in Ballester ([2006](#)) and Zhou & Chen ([2015](#)), the optimal consumption of our model is related to the matrix $(\mathbf{g}'+\mathbf{g}'^{\mathrm{T}})/2$ rather than the matrix $\mathbf{g}$ or $\mathbf{g}'$. The difference originates from the consideration of the monopolist and the pricing mechanism. Further, the difference will vanish when the matrix $\mathbf{g}$ or $\mathbf{g}'$ is symmetric according to their mathematical expressions. Thus, the intuition behind the above finding is that the monopolist's power of price discrimination is rooted in the asymmetry of the influence matrix. Note that $\mathbf{g}'-\mathbf{g}'^{\mathrm{T}}$ measures the difference in influence between each pair of customers, and then by following the interpretation provided in Candogan et al. ([2012](#)), the offered price for one customer is positively affected by how much the customer is influenced by her central peers, whereas the price is meanwhile negatively affected by the influence the customer exerts on central agents. Therefore, if one customer can influence the others much more and be influenced by the others much less, the offered price for the customer will be much more favourable, which can be interpreted as the price compensation for customers who exert influence.

## 4. Extended model and discussions

### 4.1. Solutions of extended model

As we have stated, the customer sequence always contains multiple categories rather than the two categories discussed in the basic model. Accordingly, when the customer sequence contains $k$ parts ($k \geq 2$) according to their purchasing time, we denote that these parts constitute the sequence set $T := \{t_1, t_2, \cdots, t_k\}$. Thus, the above proven **Lemma 1** directly guarantees that the following **Result 3** holds.

[**Result 3**] (**Solutions of extended model**). Given the customers' purchasing sequence set $T$, the optimal consumption vector $\mathbf{x}_T$, the optimal price vector $\mathbf{p}_T$ and the optimal profit $\pi_T$ in this extend case have the following specific expressions:

$$\mathbf{x}_T = \left( \Lambda_\beta - \frac{\mathbf{g}_T + \mathbf{g}_T^{\mathrm{T}}}{2} \right)^{-1} \left( \frac{\boldsymbol{\alpha} - c\mathbf{1}}{2} \right), \tag{19}$$

$$\mathbf{p}_T = \frac{\boldsymbol{\alpha} + c\mathbf{1}}{2} + \frac{\left( \mathbf{g}_T - \mathbf{g}_T^{\mathrm{T}} \right)}{2} \left( \Lambda_\beta - \frac{\mathbf{g}_T + \mathbf{g}_T^{\mathrm{T}}}{2} \right)^{-1} \left( \frac{\boldsymbol{\alpha} - c\mathbf{1}}{2} \right), \tag{20}$$

and

$$\pi_T = \left( \frac{\boldsymbol{\alpha} - c\mathbf{1}}{2} \right)^{\mathrm{T}} \left( \Lambda_\beta - \frac{\mathbf{g}_T + \mathbf{g}_T^{\mathrm{T}}}{2} \right)^{-1} \left( \frac{\boldsymbol{\alpha} - c\mathbf{1}}{2} \right). \tag{21}$$

Here, the relationship between $\mathbf{g}_T$ and $\mathbf{g}$ is

$$\mathbf{g}_T^{t_k t_k} = \mathbf{g}^{t_k t_k} \quad \text{and} \quad \mathbf{g}_T^{t_j t_j} = \mathbf{g}^{t_j t_j} + diag(\mathbf{g}^{t_j(\cup_{j+1 \leq i \leq k} t_i)} \mathbf{A}_T^{\cup_{j+1 \leq i \leq k} t_i} \mathbf{g}^{(\cup_{j+1 \leq i \leq k} t_i) t_j}), \quad j = k-1, k-2, \cdots, 1, \tag{22}$$

where $\mathbf{A}_T^{\cup_{j+1 \leq i \leq k} t_i} := \left( \Lambda_\beta - \mathbf{g}_T^{(\cup_{j+1 \leq i \leq k} t_i)(\cup_{j+1 \leq i \leq k} t_i)} \right)^{-1}$.

***Proof***. See **Appendix C** for details. □

**Result 3** provides the recursive form of receiving the targeted matrix $\mathbf{g}_T$, which extends the basic model. Thus far, no matter how many categories are contained by the customer sequence, the proven **Result 3** is the approach for transforming it into the equivalent *simultaneous-move game* to provide direct results.

### 4.2. Comparisons with the simultaneous-move game

In our model setting, if all the customers make their consumption decisions simultaneously, how does each customer's optimal consumption and the monopolist's optimal profit change? The following **Lemma 2** provides the answer, where two matrixes $\mathbf{x}$ and $\mathbf{y}$ satisfy $\mathbf{x} \geq \mathbf{y}$ if and only if $\mathbf{x}(i, j) \geq \mathbf{y}(i, j)$, $\forall i, j \in N$.

[**Lemma 2**] (**Comparing the customers' optimal consumption vectors and the monopolist's optimal**

**profits of two games**). Given the customers' purchasing sequence set $T$, let $\mathbf{x}_{se}$ (or $\mathbf{x}_{si}$) denote the optimal consumption vector and $\pi_{se}$ (or $\pi_{si}$) denote the highest profit of the monopolist in the *sequential-move game* (or *simultaneous-move game*). Thus, it holds that $\mathbf{x}_{se} \geq \mathbf{x}_{si}$ and $\pi_{se} \geq \pi_{si}$. In addition, when $\mathbf{g}$ is an upper or a lower triangular matrix, it holds that $\mathbf{x}_{se} = \mathbf{x}_{si}$ and $\pi_{se} = \pi_{si}$.

*Proof*. See **Appendix D** for details. □

Furthermore, from the perspective of the monopolist, the existence of leading customers will increase or at least not decrease the monopolist's profit, and thus, cultivating some leading customers will benefit the monopolist. Additionally, Lemma 2 lays a foundation to measure the value of leading customers, which is the focus of the next section.

Before discussing this value, however, we further pay attention to the offered price vectors $\mathbf{p}_{se}$ and $\mathbf{p}_{si}$ in the two kinds of games. By recalling Equations (16) and (20), the fact that $\mathbf{g} - \mathbf{g}^T = \mathbf{g}_T - \mathbf{g}_T^T$ guarantees that

$$\mathbf{p}_{se} - \mathbf{p}_{si} = \underbrace{\frac{\left(\mathbf{g} - \mathbf{g}^T\right)}{2}}_{\substack{\text{Effect from}\\\text{Matrix asymmetry}}} \underbrace{\left(\left(\Lambda_{\boldsymbol{\beta}} - \frac{\mathbf{g}_T + \mathbf{g}_T^{\ T}}{2}\right)^{-1} - \left(\Lambda_{\boldsymbol{\beta}} - \frac{\mathbf{g} + \mathbf{g}^T}{2}\right)^{-1}\right)}_{\text{Effect from adding leading consumers}} \underbrace{\left(\frac{\boldsymbol{\alpha} - c\mathbf{1}}{2}\right)}_{\substack{\text{Effect from}\\\text{inner preference}}}, \tag{23}$$

which shows that the difference between $\mathbf{p}_{se}$ and $\mathbf{p}_{si}$ originates from three effects: matrix asymmetry, leading consumers' addition and inner preference. Accordingly, the properties of the relation between $\mathbf{p}_{se}$ and $\mathbf{p}_{si}$ are obtained in **Lemma 3**.

[**Lemma 3**]. (**Comparing the offered price vector of two games**). (1) When $\mathbf{g}$ is a symmetric matrix, $\mathbf{p}_{se} = \mathbf{p}_{si}$; (2) when $\mathbf{g}$ is an upper or a lower triangular matrix, $\mathbf{p}_{se} = \mathbf{p}_{si}$.

*Proof*. The first part is obvious by recalling Equation (22), and the second part originates from the proof process of Lemma 2, where it is shown that $\mathbf{g}_T = \mathbf{g}$ if $\mathbf{g}$ is an upper or a lower triangular matrix. □

Moreover, the existence of leading customers **does not** guarantee that $\mathbf{p}_{se} \geq \mathbf{p}_{si}$ or $\mathbf{p}_{se} < \mathbf{p}_{si}$. However, in most cases, adding or cultivating the leading customers will increase the monopolist's profit, which can be paid special attention to by introducing the referral mechanism.

### 4.3. Discussion

From **Lemma 1** and **Result 3**, we find that only the diagonal elements of the leading customers increase compared to the original matrix $\mathbf{g}$. Accordingly, we can also understand the effect of leading customers from the perspective of $\beta_i$ by noting that $\left(\Lambda_{\boldsymbol{\beta}} - (\mathbf{g}' + \mathbf{g}'^T)/2\right)^{-1}$ or $\left(\Lambda_{\boldsymbol{\beta}} - (\mathbf{g}_T + \mathbf{g}_T^{\ T})/2\right)^{-1}$ always appear in the result as a whole. In other words, a leading customer $i$ with his/her preference parameter $\beta_i$ in the *sequential-move game* is equivalent to a customer with a new preference parameter $\beta_i' \le \beta_i$ in the *simultaneous-move game*. Namely, the advantage of a leading position in the *sequential-move game* can be transferred to be a decrease in the preference parameter $\beta_i$ in the simultaneous-move game. Recalling the adopted utility function in Equation (1), $\beta_i$ measures the decreasing effect of marginal utility so that a smaller value means a lower decreasing effect.

# 5. The value of leading customers

This section aims to utilize the above proven lemmas and results to provide an applicable network-based referral mechanism. Because **Lemma 2** has demonstrated that adding or cultivating some leading customers will increase the monopolist's profit in most cases, we can accordingly define the value of leading customers from the perspective of increased profit. Next, recalling the designed referral network in Equation (1), we can use the defined value of leading customers to reward the leading customers to encourage them to form the referral network. Based on this basic idea, we further explore the properties and present the potential application of the provided network-based referral mechanism.

## 5.1. Definition and example

The monotonicity of the monopolist's profit helps us measure the value of the leading customers, and then **Definition 1** provides the formal expressions.

[**Definition 1**] Given the customer sequence $T \coloneqq \{t_1, t_2, \cdots, t_j, t_{j+1}, \cdots, t_k\}$, we denote $T(j) \coloneqq \{t_1, t_2, \cdots, t_j, (t_{j+1} \cup t_{j+2} \cup \cdots \cup t_k)\}$ and define $T(0) \coloneqq T$; then the value of leading customer set $t_j$ is denoted as $value(t_j)$ with the following expression:

$$value(t_j) \coloneqq \pi_{se}(\mathbf{g}_{T(j)}) - \pi_{se}(\mathbf{g}_{T(j-1)}), \tag{24}$$

where the profit function $\pi_{se}(\cdot)$ has been shown in Equation (21). □

Next, two basic network structures are presented as examples to provide an intuitive understanding of the definition and to further illustrate some interesting findings.

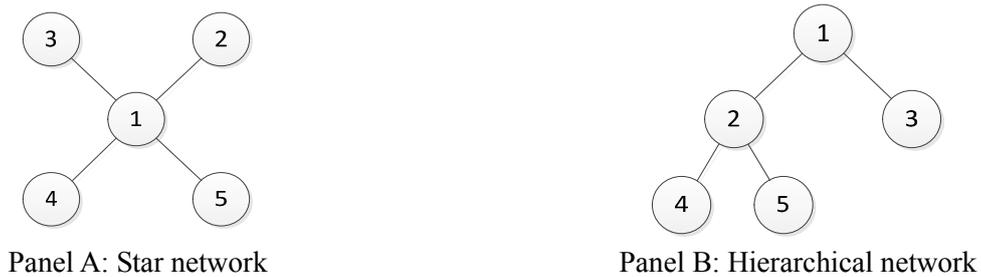

Panel A: Star network                    Panel B: Hierarchical network

**Fig. 1**. Two basic network structures

As shown in **Fig. 1**, the adopted two basic networks contain the same number of nodes and links, which can be regarded as two typical basic structures of a complex network. According to **Definition 1**, **Table 1** displays the values of leading customer sets under different customer sequences.

**Table 1**. The values of leading customer sets under different customer sequences.

| Case No. | customer sequence | star network | | hierarchical network | |
|---|---|---|---|---|---|
| | | value of leading customer | profit | value of leading customer | profit |
| 1 | {1,2,3,4,5} | ---- | 0.3929 | ---- | 0.3800 |
| 2 | {1},{2,3,4,5} | value({1})=0.0453 | 0.4382 | value({1})=0.0124 | 0.3924 |
| 3 | {2},{1,3,4,5} | value({2})=0.0048 | 0.3977 | value({2})=0.0249 | 0.4049 |
| 4 | {2},{1,3},{4,5} | value({2})=0.0048; value({1,3})=0.0212 | 0.4189 | value({2})=0.0249; value({1,3})=0 | 0.4049 |
| 5 | {2},{4,5},{1,3} | value({2})=0.0048; value({4,5})=0.0090 | 0.4067 | value({2})=0.0249; value({4,5})=0 | 0.4049 |
| 6 | {1,2},{3,4,5} | value({1,2})=0.0321 | 0.4250 | value({1,2})=0.0214 | 0.4014 |

**Note**. In this example, $\alpha_i = 1$, $\beta_i = 5$ and $c_i = 0$, for any $i \in N$.

By comparing these results, the following findings can be summarized. Although two networks share the same number of nodes and links, the profit of the monopolist is different according to case No. 1. Accordingly, the network structure in itself can influence profits, and thus, some network structures are preferred by the monopolist. In addition, by comparing case No. 2 and No. 3, choosing different leading customers will lead to different profits, and thus, the strategy of choosing the leading customer set is meaningful for the monopolist in increasing the achieved profit. Moreover, the difference between case No. 4 and No. 5 illustrates that the order of the customer sequence will be a factor that influences the profit of the monopolist. Accordingly, when the customer sequence comprises more than two sets, it will be an interesting problem to determine the optimal customer sequence to achieve the largest profit. Last but not least, it must not increase the profit that more nodes are chosen as the leading customers by comparing cases No. 6 and No. 2. Specifically, compared with case No. 2, the profit decreases when adding node 2 into the leading customer set in the star network, whereas the hierarchical network shows the opposite. Accordingly, the relationship between the achieved profit and the amount of leading customers is not monotonous.

*5.2. Network-based referral strategy*

Aside from the above interesting findings, this subsection focuses on coping with exactly how many rewards should be allocated to the leading customers if they benefit enlarging the purchase of the targeted product, especially in the crowdfunding-based marketing pattern. Note that the approach of allocating rewards along the referral network is the core of providing the referral mechanism. Specifically, as an advertising strategy, the monopolist enables rewarding the existing customers who successfully recommend some newcomers to participate in the crowdfunding. One direct problem is exactly how much of a reward should be allocated. As we have explained, the defined value of leading customers can be a tool to cope with the problem. On the one hand, when a newcomer participates in the crowdfunding, the values of the leading customers are likely to change, which can quantitatively reflect the effect of a successful recommendation. On the other hand, the change in the values of the leading customers originates from the change in the monopolist's profits, so it is helpful for the mechanism design of marketing, especially when social learning becomes more and more important with the development of modern crowdfunding.

Before going into details, we first answer a fundamental question: will adding a new customer increase or at least not decrease all the leading customers' values no matter what their sequence positions are? If the answer is YES, we can argue that each leading customer with a direct or indirect influence on the newcomer will benefit from the arrival of a new customer. **Lemma 4** provides the answer.

[**Lemma 4**] Given the customer sequence $T := \{t_1, t_2, \cdots, t_j, t_{j+1}, \cdots, t_k\}$ and the comprehensive network **g** shown in Equation (1b), a newcomer denoted as the $t_{k+1}$ is added into the customer sequence and then the new customer sequence is denoted as $T_{+1} := \{t_1, t_2, \cdots, t_j, t_{j+1}, \cdots, t_k, t_{k+1}\}$. As a result, $value(t_j | T_{+1}) \geq value(t_j | T)$ holds for any $j \in \{1, 2, \cdots, k\}$, where $value(t_i | T_{+1})$ and $value(t_i | T)$ represent the value of leading customer set $t_j$ in the customer sequence $T_{+1}$ and $T$, respectively.

***Proof***. See **Appendix E** for details.                                                    □

**Lemma 4** actually points out that when a newcomer is added at the end of the customer sequence, all the leading customers of the newcomer will not decrease their leading values. Based on the theoretic basis of **Lemma 4**, Equation (24) further guides us to measure the newcomer's contributions on increasing each leading customer's value. To clarify, one example is provided below, where Panel A of Fig. 2 displays a simple comprehensive network comprising four customers with their customer sequence $\{\{1\}, \{2\}, \{3\}, \{4\}\}$ and Panel B of Fig. 2 displays a new customer's arrival and their customer sequence is changed into $\{\{1\}, \{2\}, \{3\}, \{4\}, \{5\}\}$. Next, given that all the customers share the identical preference parameters, each leading customer's value and their changes in adding a new comer are listed in **Table 2**.

**Table 2** validates the result of **Lemma 4**, where all the leading customer's value increases. As displayed in **Fig. 2**, new customer No. 5 is recommended by customer No. 4, so that the pair have direct influence. Furthermore, the leading customers at different sequence positions truly increase at different values; the value directly influenced by the newcomer has the largest increase and is called **the rule of "closeness"**, and an earlier leading customer has a lower value increase and is called **the rule of "decreasing"**. By recalling the background provided in the Introduction, another finding emerging from this example is that not only does the one who directly recommends the new comer benefit from the newcomer's arrival, but the earlier leading customers also benefit, although they only have indirect influence.

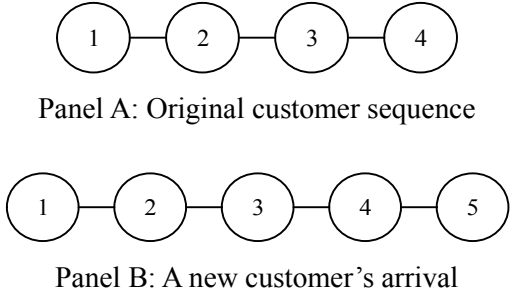

Panel A: Original customer sequence

Panel B: A new customer's arrival

**Fig. 2**. Two customer sequences.

**Table 2**. Each leading customer's value and the change.

| Panel A | Leading value | Panel B | Leading value | Increased value |
|---------|---------------|---------|---------------|-----------------|
| {1} | 0.0038 | {1} | 0.0038 | +0.0000 |
| {2} | 0.0057 | {2} | 0.0058 | +0.0001 |
| {3} | 0.0056 | {3} | 0.0063 | +0.0007 |
| {4} | 0 | {4} | 0.0057 | +0.0057 |
| ---- | ---- | {5} | 0 | ---- |

**Note**. $\alpha_i = 1$, $\beta_i = 5$, $c_i = 0$ and $\eta = 1$ for any $i \in N$.

Inspired by this simple example, we further pay close attention to the general rule of the proposed network-based referral strategy. Thus far, we know that adding a newcomer will increase the leading values of these customers along the referral line and we also present the approach of calculating the increased leading values. Next, we illustrate the two rules discovered in the above simple example to facilitate the crowdfunding initiator to reward the related leading customers owing to their successful recommendations. Furthermore, our model only allows that one newcomer accepts one existing customer's recommendation, although the newcomer can actually receive several customers' information; in other words, each customer has at most one referrer. The above-mentioned mechanism of acceptance will ease the allocations of rewards because all the related leading customers constitute a line network. Deepening the analysis of the above simple example, **Property 1** provides the general rules for allocating the increased profits from the newcomer in our proposed network-based referral strategy.

[**Property 1**] (**General rules of proposed network-based referral strategy**). The rule of "closeness" and the rule of "decreasing" should be adopted when allocating the increased profits from the newcomer along the line network consisting of the related leading customers, given that all the customers share the identical preference parameters.

*Proof*. See **Appendix F** for details. □

To sum up, customers can be encouraged by obtaining rewards when they successfully recommend newcomers to participate. Especially with the development of the sharing economy, for example, as in a crowdfunding-based marketing pattern, it becomes more convenient and even important for customers to utilize their social networks to recommend the targeted product. This subsection provides an alternative approach to how to encourage customers who make successful recommendations, as well as of determining exactly how many profits should be allocated. More precisely, the two rules demonstrated in **Property 1** can qualitatively guide the monopolist in how to allocate the additional profits to each leading customer when a new customer comes into the market. Specifically, the customer who directly recommends and affects the newcomer will be allocated most of the profits because this customer contributes most to the new customer's entrance into the market, and the earlier customer in the line network should be allocated much less, given the longer distance to the newcomer. In addition, the proven **Lemma 2** provides a quantitative approach to coping with how much profit should actually be returned based on the calculated values of leading customers.

### 5.3. Discussion

Recalling the utility function defined in Equation (1a), there are two effects discussed and considered. One is the influence of the referrers and the other is the influence of the successfully recommended followers. In the real market, modern technology allows us to record the information of successful recommendation, and thus, the referral network $R$ is clear, especially in the online crowdfunding-based marketing platform. The remaining problem is how the strength parameter $\eta$ affects the strategy of allocation, and Property 2 provides the answer.

[**Property 2**]. (**The effect of strength parameter** $\eta$). When $\eta = 0$, each leading customer's value equals 0 in our model setting. In addition, each leading customer's value increases with the rise of $\eta$.

*Proof*. When $\eta = 0$, the comprehensive matrix $\mathbf{g}$ becomes a lower triangular matrix, and thus, Lemma 2 guarantees that each leading customer's value equals 0 in our model setting. In addition, Equations (21) and (22) guarantee that a larger $\eta$ means larger diagonal elements in $\mathbf{g}_T$ and will thus cause a larger value of each leading customer. □

**Property 2** points out that if the leading customers are not affected by their followers' consumption decisions (i.e., $\eta = 0$), the leading values vanish or say the leading positions do not take effect. However, if the leading customers are greatly influenced by their followers' consumption decisions (i.e., a larger $\eta$), the leading positions become much more valuable. In a real market, when some customers accept the recommendation, the referrer truly increases his/her utility because some peers agree with the targeted project, as the referrer has, and meanwhile much larger amounts consumed by the followers means a higher increase in utility. Thus, $\eta$ should be larger than 0 in a real market in most cases. Following the result of **Property 2**, if we can obtain each customer's information about his/her reaction to their peers' consumption decision in advance, the ones with a much greater reaction should be chosen as the leading ones from the perspective of $\eta$ when the other influence factors prove identical.

## 6. Conclusions

Currently, crowdfunding is gradually becoming a modern marketing mode via the Internet. Facing this trend, we aim to provide feasible marketing strategies to promote modern crowdfunding-based crowdfunding patterns. To this end, we first analysed the features of modern crowdfunding-based marketing patterns and find the following three features: monopoly structure, network externalities and customer sequence. Building on the work of Candogan et al. (2012), Zhou & Chen (2015) and Zhou & Chen (2017), we defined the value of leading customers by solving the game models (i.e., basic model and extended model) contained within the above three features. Based on the solution of a basic model comprising only leaders and followers, the relationship between the *simultaneous-move game* and *sequence-move game* is revealed, and then the extended model whose customer sequence consists of more than two sets is further solved. According to the defined leading customer's values, we further focus on providing a network-based referral mechanism in the crowdfunding-based market pattern by exploring how to allocate rewards along the customer sequence according to their leading values and by uncovering the rules and properties of the proposed network-based referral mechanism.

The main marketing strategies for promoting crowdfunding-based marketing patterns are summarized as follows. When a customer sequence with more than two customer sets is actually formed, the crowdfunding initiator can also earn greater profits by rewarding the leading customers for utilizing their social networks for recommendation. The demonstrated strategies for allocating the rewards have the following rules: (1) the customer who directly recommends the newcomer will be allocated the most additional profits and (2) the earlier leading customer in the line network should be allocated fewer profits, or exactly exponentially decreasing profits. We tend to believe that the proposed quantitative strategies are easy to operate in a real market.

**Acknowledgement:** This work was supported by the National Natural Science Foundation of China under Grant 71501034 and 71771041, and by the China Postdoctoral Science Foundation under Grant 2016M590230 and 2017T100183.

**Appendix A**. (Proof of **Result 1**).

By following the three solution steps mentioned in Section 3.1, the result of ***Step 1***, namely, the specific expression of Equation (4), is $\mathbf{x}^b = [\Lambda_\beta^b - \mathbf{g}^{bb}]^{-1}[\boldsymbol{\alpha}^b - \mathbf{p}^b + \mathbf{g}^{ba}\mathbf{x}^a]$. Similarly, the result of ***Step 2*** is $\mathbf{x}^a = \left[\Lambda_\beta^a - \mathbf{g}^{aa} - \mathbf{g}^{ab}\mathbf{A}^b\mathbf{g}^{ba} - diag(\mathbf{g}^{ab}\mathbf{A}^b\mathbf{g}^{ba})\right]^{-1}\left(\boldsymbol{\alpha}^a - \mathbf{p}^a + \mathbf{g}^{ab}\mathbf{A}^b(\boldsymbol{\alpha}^b - \mathbf{p}^b)\right)$.

Next, substituting $\mathbf{x}^a$ into $\mathbf{x}^b$, we obtain $\mathbf{x}^b = (\mathbf{A}^b\mathbf{g}^{ba}\mathbf{B}_{se}^a)(\boldsymbol{\alpha}^a - \mathbf{p}^a) + (\mathbf{A}^b + \mathbf{A}^b\mathbf{g}^{ba}\mathbf{B}_{se}^a\mathbf{g}^{ab}\mathbf{A}^b)(\boldsymbol{\alpha}^b - \mathbf{p}^b)$, where $\mathbf{B}_{se}^a := \left[\Lambda_\beta^a - \mathbf{g}^{aa} - \mathbf{g}^{ab}\mathbf{A}^b\mathbf{g}^{ba} - diag(\mathbf{g}^{ab}\mathbf{A}^b\mathbf{g}^{ba})\right]^{-1}$. Thus far, the optimal consumptions $\mathbf{x}^a$ and $\mathbf{x}^b$ of two categories are both expressed as the function of the offered price vector $\mathbf{p}$, and then we further express them in matrix form:

$$\begin{pmatrix} \mathbf{x}^a \\ \mathbf{x}^b \end{pmatrix} = \mathbf{M}\begin{pmatrix} \boldsymbol{\alpha}^a - \mathbf{p}^a \\ \boldsymbol{\alpha}^b - \mathbf{p}^b \end{pmatrix},$$

where the matrix $\mathbf{M}$ takes the following form:

$$\mathbf{M} = \begin{bmatrix} \mathbf{B}_{se}^a & \mathbf{B}_{se}^a\mathbf{g}^{ab}\mathbf{A}^b \\ \mathbf{A}^b\mathbf{g}^{ba}\mathbf{B}_{se}^a & \mathbf{A}^b + \mathbf{A}^b\mathbf{g}^{ba}\mathbf{B}_{se}^a\mathbf{g}^{ab}\mathbf{A}^b \end{bmatrix}.$$

Lastly, ***Step 3*** is conducted to achieve the solutions of the *basic model*. Specifically, the optimal consumption of the customers is

$$\mathbf{x}_{se} = [\mathbf{M}^{-1} + \mathbf{M}^{-\mathrm{T}}]^{-1}(\boldsymbol{\alpha} - c\mathbf{1}),$$

the optimal offered pricing vector is

$$\mathbf{p}_{se} = \boldsymbol{\alpha} - \mathbf{M}^{-1}[\mathbf{M}^{-1} + \mathbf{M}^{-\mathrm{T}}]^{-1}(\boldsymbol{\alpha} - c\mathbf{1}),$$

and, accordingly, the largest profit gained by the monopolist is

$$\pi_{se} = \frac{(\boldsymbol{\alpha} - c\mathbf{1})^{\mathrm{T}}[\mathbf{M}^{-1} + \mathbf{M}^{-\mathrm{T}}]^{-1}(\boldsymbol{\alpha} - c\mathbf{1})}{2}.$$

**Appendix B**. (Proof of **Lemma 1**).

Replacing $\mathbf{g}$ by $\mathbf{g}'$, the consumer's utility function in the *simultaneous-move game* is

$$u_i(x_i; \mathbf{x}_{-i}, p_i) = \alpha_i x_i - \frac{1}{2}\beta_i x_i^2 + \sum_{j=1}^{N} g_{ij}' x_i x_j - p_i x_i.$$

According to **Result 2** shown in Section 3.3, we immediately obtain the expressions of the optimal consumption vector and the optimal price vector by replacing $\mathbf{g}$ with $\mathbf{g}'$, respectively. Thus, $\mathbf{B}_{si}^a$ is further expressed as $\mathbf{B}_{si}^a(\mathbf{g}')$ with the following formula:

$$\mathbf{B}_{si}^a(\mathbf{g}') = \left[\Lambda_\beta^a - \mathbf{g}^{aa}{}' - \mathbf{g}^{ab}\mathbf{A}^b\mathbf{g}^{ba}\right]^{-1},$$

because the matrix transformation does not change $\mathbf{g}^{ab}$, $\mathbf{g}^{ba}$, $\mathbf{g}^{bb}$ as well as the corresponding $\mathbf{A}^b$. Moreover, according to the condition that $\mathbf{g}^{aa}{}' = \mathbf{g}^{aa} + diag(\mathbf{g}^{ab}\mathbf{A}^b\mathbf{g}^{ba})$ and the expression of $\mathbf{B}^a_{se}$, it immediately holds that

$$\mathbf{B}^a_{si}(\mathbf{g}') = \mathbf{B}^a_{se}(\mathbf{g}),$$

where $\mathbf{B}^a_{se}$ is highlighted as the function of the matrix $\mathbf{g}$. Next, the above Equation guarantees that

$$\mathbf{M}(\mathbf{g}) = \mathbf{A}(\mathbf{g}'),$$

where the matrix $\mathbf{M}$ and $\mathbf{A}$ are highlighted as the function of $\mathbf{g}$ and $\mathbf{g}'$, respectively. Thus, by recalling the expressions of $\mathbf{x}_{se}$, $\mathbf{p}_{se}$, $\mathbf{x}_{si}$, $\mathbf{p}_{si}$, $\pi_{si}$ and $\pi_{se}$, **Lemma 1** holds.

**Appendix C**. (Proof of **Result 3**).

The part $t_k$ has no followers, so their decisions are not affected by the followers. As a result, the transformed matrix of $t_k$ remains unchanged; specifically, $\mathbf{g}_T^{t_k t_k} = \mathbf{g}^{t_k t_k}$. Then, the part $t_{k-1}$ has the follower part $t_k$ and therefore **Lemma 1** guarantees that

$$\mathbf{g}_T^{t_{k-1} t_{k-1}} = \mathbf{g}^{t_{k-1} t_{k-1}} + diag(\mathbf{g}^{t_{k-1} t_k}\mathbf{A}_T^{t_k}\mathbf{g}^{t_k t_{k-1}}),$$

which accords with **Result 1** when only two categories exist.

Next, when the part $t_{k-2}$ is considered, it has two parts of followers: $t_k$ and $t_{k-1}$, which have an influence on $t_k$. Then, the combined two parts' new influence matrix $\mathbf{g}_T^{(t_{k-1} \cup t_k)(t_{k-1} \cup t_k)}$ can be reached as below

$$\mathbf{g}_T^{(t_{k-1} \cup t_k)(t_{k-1} \cup t_k)} = \begin{bmatrix} \mathbf{g}_T^{t_{k-1} t_{k-1}} & \mathbf{g}_T^{t_{k-1} t_k} \\ \mathbf{g}_T^{t_k t_{k-1}} & \mathbf{g}_T^{t_k t_k} \end{bmatrix} = \begin{bmatrix} \mathbf{g}_T^{t_{k-1} t_{k-1}} & \mathbf{g}^{t_{k-1} t_k} \\ \mathbf{g}^{t_k t_{k-1}} & \mathbf{g}^{t_k t_k} \end{bmatrix},$$

where only the part $\mathbf{g}_T^{t_{k-1} t_{k-1}}$ is different from the original $\mathbf{g}^{(t_{k-1} \cup t_k)(t_{k-1} \cup t_k)}$. Accordingly, once the two parts are considered together, **Lemma 1** is also useful for achieving the transformed $\mathbf{g}^{t_{k-2} t_{k-2}}$:

$$\mathbf{g}_T^{t_{k-2} t_{k-2}} = \mathbf{g}^{t_{k-2} t_{k-2}} + diag(\mathbf{g}^{t_{k-2} (t_{k-1} \cup t_k)}\mathbf{A}_T^{t_{k-1} \cup t_k}\mathbf{g}^{(t_{k-1} \cup t_k) t_{k-2}}).$$

Repeating the above process, we can achieve the transformed $\mathbf{g}^{t_j t_j}$ ($1 \le j \le k-3$) by considering the followers $t_{j+1}$, $t_{j+2}$, $\cdots$, $t_k$ as one whole part. Note that the whole part's influence matrix has been transformed as $\mathbf{g}_T^{(\cup_{j+1 \le i \le k} t_i)(\cup_{j+1 \le i \le k} t_i)}$, where $\mathbf{g}_T^{t_i t_i}$ ($j+1 \le i \le k$) has been achieved in the above processes. As a result, **Lemma 1** guarantees $\mathbf{g}_T^{t_j t_j} = \mathbf{g}^{t_j t_j} + diag(\mathbf{g}^{t_j(\cup_{j+1 \le i \le k} t_i)}\mathbf{A}_T^{\cup_{j+1 \le i \le k} t_i}\mathbf{g}^{(\cup_{j+1 \le i \le k} t_i) t_j})$. In all, **Result 3** holds.

**Appendix D**. (Proof of **Lemma 2**).

We first prove this lemma when all the customers are only divided into two categories: leaders and followers. In this case, it first holds that

$$\left(\Lambda_{\beta}-\frac{\mathbf{g}'+\mathbf{g}'^{\mathrm{T}}}{2}\right)^{-1}-\left(\Lambda_{\beta}-\frac{\mathbf{g}+\mathbf{g}^{\mathrm{T}}}{2}\right)^{-1}=\left(\Lambda_{\beta}-\frac{\mathbf{g}+\mathbf{g}^{\mathrm{T}}}{2}\right)^{-1}\left(\frac{\mathbf{g}'+\mathbf{g}'^{\mathrm{T}}}{2}-\frac{\mathbf{g}+\mathbf{g}^{\mathrm{T}}}{2}\right)\left(\Lambda_{\beta}-\frac{\mathbf{g}'+\mathbf{g}'^{\mathrm{T}}}{2}\right)^{-1},$$

and then, according to the relationship between $\mathbf{g}$ and $\mathbf{g}'$, we have

$$\left(\frac{\mathbf{g}'+\mathbf{g}'^{\mathrm{T}}}{2}-\frac{\mathbf{g}+\mathbf{g}^{\mathrm{T}}}{2}\right)=diag\left(\begin{matrix}\mathbf{g}^{ab}\mathbf{A}^{b}\mathbf{g}^{ba}&\\&\mathbf{0}_{(N-m)\times(N-m)}\end{matrix}\right),$$

where $\mathbf{0}_{(N-m)\times(N-m)}$ is an $(N-m)\times(N-m)$ matrix. Next, by recalling Assumption 1, all the elements are not less than 0 in the following three matrixes:

$$\left(\Lambda_{\beta}-\frac{\mathbf{g}+\mathbf{g}^{\mathrm{T}}}{2}\right)^{-1},\ \ diag\left(\begin{matrix}\mathbf{g}^{ab}\mathbf{A}^{b}\mathbf{g}^{ba}&\\&\mathbf{0}_{(N-m)\times(N-m)}\end{matrix}\right),\text{ and }\left(\Lambda_{\beta}-\frac{\mathbf{g}'+\mathbf{g}'^{\mathrm{T}}}{2}\right)^{-1}.$$

As a result, it immediately holds that

$$\left(\Lambda_{\beta}-\frac{\mathbf{g}'+\mathbf{g}'^{\mathrm{T}}}{2}\right)^{-1}\geq\left(\Lambda_{\beta}-\frac{\mathbf{g}+\mathbf{g}^{\mathrm{T}}}{2}\right)^{-1}.$$

Thus, note that no matter how many categories are contained in the customer sequence, the above result holds for any two divisions contained in sub-matrixes. In all, **Lemma 2** holds.

**Appendix E**. (Proof of **Lemma 4**)

According to **Definition 1**, we immediately achieve the following equation:

$$value(t_j\,|\,T_{+1})=\left(\frac{\boldsymbol{\alpha}-c\mathbf{1}}{2}\right)^{\mathrm{T}}\left(\left(\Lambda_{\beta}-\frac{\mathbf{g}_{T_{+1}(j)}+\mathbf{g}_{T_{+1}(j)}{}^{\mathrm{T}}}{2}\right)^{-1}-\left(\Lambda_{\beta}-\frac{\mathbf{g}_{T_{+1}(j-1)}+\mathbf{g}_{T_{+1}(j-1)}{}^{\mathrm{T}}}{2}\right)^{-1}\right)\left(\frac{\boldsymbol{\alpha}-c\mathbf{1}}{2}\right)$$

$$=\left(\frac{\boldsymbol{\alpha}-c\mathbf{1}}{2}\right)^{\mathrm{T}}\left(\Lambda_{\beta}-\frac{\mathbf{g}_{T_{+1}(j-1)}+\mathbf{g}_{T_{+1}(j-1)}{}^{\mathrm{T}}}{2}\right)^{-1}\left(\frac{\mathbf{g}_{T_{+1}(j)}+\mathbf{g}_{T_{+1}(j)}{}^{\mathrm{T}}}{2}-\frac{\mathbf{g}_{T_{+1}(j-1)}+\mathbf{g}_{T_{+1}(j-1)}{}^{\mathrm{T}}}{2}\right)\left(\Lambda_{\beta}-\frac{\mathbf{g}_{T_{+1}(j)}+\mathbf{g}_{T_{+1}(j)}{}^{\mathrm{T}}}{2}\right)^{-1}\left(\frac{\boldsymbol{\alpha}-c\mathbf{1}}{2}\right),$$

Next, we add an isolated node into the given customer sequence $T$, and therefore we also obtain a new customer sequence denoted as $T_{+1}'$ with $k+1$ elements. Note that the added node is isolated so that the new formed influence matrix $\mathbf{g}'$ has the following expressions: $g_{ij}'=g_{ij}$, $g_{i(n+1)}'=0$ and $g_{(n+1)j}'=0$ for any $i,j\in\{1,2,\cdots,n\}$, and $g_{(n+1)(n+1)}'=0$.

Thus, recalling **Result 3** proven above, we immediately achieve the following two results:

(i) All the elements in $\mathbf{g}_{T_{+1}(j)}$ are no less than $\mathbf{g}_{T_{+1}(j)}'$ for any $j\in\{1,2,\cdots,k\}$, which is denoted as $\mathbf{g}_{T_{+1}(j)}\geq\mathbf{g}_{T_{+1}(j)}'$, and therefore it holds that

$$\left(\Lambda_{\beta}-\frac{\mathbf{g}_{T_{+1}(j)}+\mathbf{g}_{T_{+1}(j)}{}^{\mathrm{T}}}{2}\right)^{-1}\geq\left(\Lambda_{\beta}-\frac{\mathbf{g}_{T_{+1}(j)}'+\mathbf{g}_{T_{+1}(j)}'^{\mathrm{T}}}{2}\right)^{-1};$$

(ii) The series of equations provided in Equation (22) guarantee that

$$\frac{\mathbf{g}_{T_{+1}(j)} + \mathbf{g}_{T_{+1}(j)}^{\mathrm{T}}}{2} - \frac{\mathbf{g}_{T_{+1}(j-1)} + \mathbf{g}_{T_{+1}(j-1)}^{\mathrm{T}}}{2} \geq \frac{\mathbf{g}_{T_{+1}(j)}' + \mathbf{g}_{T_{+1}(j)}'^{\mathrm{T}}}{2} - \frac{\mathbf{g}_{T_{+1}(j-1)}' + \mathbf{g}_{T_{+1}(j-1)}'^{\mathrm{T}}}{2} \geq \mathbf{0}_{(n+1)\times(n+1)}.$$

By considering the above results together, we further obtain $value(t_j \mid T_{+1}) \geq value(t_j \mid T_{+1}')$. Lastly, because the added node is isolated, it does not change the original result under the given customer sequence $T$. As a result, $value(t_j \mid T_{+1}) \geq value(t_j \mid T)$ holds.

**Appendix F**. (Proof of **Property 1**).

Without loss of generality, a line network contains $n$ customers, and their IDs are 1 to $n$, which also reflects their sequence orders, as displayed in **Fig. 2**. Given $\alpha$ and $\beta$, when a newcomer $n+1$ is added into the line network, we next calculate how much value is increased of the existing $n$ leading customers. For any $j \in \{1, 2, \cdots, n\}$, we denote $\Delta value(t_j) = value(t_j \mid T_{+1}) - value(t_j \mid T)$. Thus, according to Equation (22) in **Result 3** and the fact that the line network structure can be expressed as a tridiagonal matrix, we can achieve the following relations:

$$\Delta value(t_j) \propto \beta^{-(n+2-j)},$$

which implies that when $k \in \{2, \cdots, n\}$,

$$\frac{\Delta value(t_k)}{\Delta value(t_{k-1})} \propto \beta.$$

Thus, the two rules hold under the line network structure and the given precondition.